# Photonic Spiking Graph Neural Network for Energy-Efficient Structured Data Processing

Wanting Yu[a], Shuiying Xiang[a,b*], Xingxing Guo[a,b*], Shangxuan Shi[a], Haowen Zhao[a], Xintao Zeng[a], Yahui Zhang[a], Hongbo Jiang[a], and Yue Hao[b]

[a]State Key Laboratory of Integrated Service Networks, Xidian University, Xi'an 710071, China.
[b]State Key Discipline Laboratory of Wide Bandgap Semiconductor Technology, School of Microelectronics, Xidian University, Xi'an 710071, China.
*Email: syxiang@xidian.edu.cn, xxguo@xidian.edu.cn

**ABSTRACT:** Photonic computing exhibits broad application potential in signal processing and artificial intelligence (AI) acceleration owing to its high computational speed, low energy consumption, and intrinsic parallelism. Existing photonic computing research has predominantly concentrated on convolutional neural networks (CNNs) and fully connected neural networks (FCNNs), which are typically applied to tasks such as image classification and object detection. However, these traditional network architectures face inherent limitations in effectively processing graph-structured data. Graph neural networks (GNNs), specifically designed for modeling and analyzing graph-structured data, effectively capture the complex relationships among data points. We propose an computational architecture termed the photonic spiking graph neural network (PSGNN). The proposed architecture integrates the efficient structural modeling capability of GNNs, the temporal dynamics of spiking neurons, and the ultra-high speed and low power consumption advantages of photonic chips for parallel computation. Through hardware–software co-optimization, a bias-term simulation mechanism tailed for photonic computing chip is implemented via a feature-dimension expansion strategy, thereby facilitating effective model training. Experimental results on the KarateClub (PubMed) datasets demonstrate a training accuracy of 100% (92 ± 2%) and a test accuracy of 97% (90 ± 1%). A hardware experimental platform based on a silicon-photonics-based 4 × 4 Mach–Zehnder interferometer (MZI) array was constructed to perform classification experiments on the KarateClub (PubMed) datasets, achieving a test accuracy of 93%, thereby validating the effectiveness of the proposed architecture in structured-data processing tasks. The proposed system achieved an inference latency of 97 ps, demonstrating remarkable acceleration within the optical domain. The photonic linear computation achieved an energy efficiency of 280 GOPS/W and a computational density of 52 GOPS/mm². These results underscore the potential of PSGNN for deployment across multiple domains, including dynamic topology analysis of social networks, intelligent traffic scheduling and early-warning systems, and fault diagnosis in industrial IoT.

## 1 Introduction

With the rapid advancement of large-model technologies represented by generative pre-trained transformers (GPT) and pathways language models (PaLM), training these models on massive datasets has led to an exponential surge in the demand for AI computing power. However, traditional electronic chip architectures, constrained by the inherent bottlenecks of the von Neumann architecture, encounter challenges such as transmission latency, high energy consumption, and low computational efficiency[1-2]. In contrast, photonic computing exhibits significant potential for applications in signal processing and AI acceleration owing to its high computational speed, low power consumption, and intrinsic parallelism[3-4].

Current research on photonic computing chips primarily focuses on convolutional neural networks (CNNs) and fully connected neural networks (FCNNs), which are mainly applied to tasks such as image classification and object detection. To validate model performance, standard datasets such as MNIST[5-7], Fashion-MNIST[8-10], and ImageNet[5,11] are commonly used. Training and testing on these datasets verify the inference speed, energy efficiency, and accuracy of models powered by photonic computing chips, thereby laying the foundation for their deployment in more complex real-world scenarios. For instance, in 2023, Yuan *et al.* proposed the dual-neuron optical–artificial learning (DANTE) architecture[5], which addresses the challenges of high computational cost, substantial memory consumption, and convergence difficulties in training large-scale convolutional optoelectronic neural networks (ONNs) caused by optical diffraction modeling. Through global artificial learning and local optical learning, its effectiveness was validated on tasks including CIFAR-10 (84.91% accuracy with a 3-layer ONN), ImageNet-32 (44.26% Top-1 accuracy and 68.61% Top-5 accuracy with a 10-layer ONN), and physical experiments on MNIST (96% accuracy). Meng *et al.* developed a compact optical convolutional processing unit based on multimode interference (MMI)[6], which integrates wavelength-division multiplexing to achieve parallel computation of three 2×2 convolutional kernels, yielding a classification accuracy of 92.17% on MNIST. In 2024, Zhang *et al.* proposed a matrix core chip based on photonic crystal nanobeam cavity (PCNC)[7]. Serving as the computational core for CNNs, this chip accomplished cat-dog recognition (85.9% accuracy) using a 3×3 array configuration and handwritten digit recognition (MNIST, 87.0% accuracy) with a six-channel chip. Huang *et al.* proposed a compact multilayer optical neural network (MONN) architecture [12], which employs two optical masks to perform dual convolutional operations, with a quantum-dot thin film inserted as an all-optical nonlinear activation layer emulating the ReLU function. This architecture successfully performed multiple visual tasks: achieving 77.0% accuracy with a single core and 80.0% with multiple cores in the 10-class Quick Draw hand-drawn object classification, and 78.1% accuracy with a single core and 83.0% with multiple cores in the 5-class Weizmann human action classification. Cheng *et al.* proposed a trainable diffractive optical neural network (TDONN) [13] with in-situ training capability. This network employs a fully connected architecture, modulates diffractive unit parameters via metallic microheaters, and incorporates a customized stochastic gradient descent algorithm. It supports in-situ training in the optical domain, yielding inference results after a single optical forward pass. Successful inference was achieved on the MNIST visual dataset (86% accuracy), the Spoken_numbers_pcm audio dataset (82% accuracy), and tactile data collected by a haptic glove for four-class classification (82% accuracy). Afifi *et al.* proposed two silicon photonics-based hardware accelerators[14] for accelerating Transformer neural networks in large language models (LLMs) and graph neural networks (GNNs) in graph data processing. In 2025, Sun *et al.* proposed and experimentally validated an efficient photonic convolutional accelerator based on lossless pattern segmentation fan-in[15], achieving 95.2% classification accuracy on MNIST and 87.9% on Fashion-MNIST. Wu et al. proposed a silicon quantum photonic circuit scheme[16] consisting of two convolutional layers and two fully connected layers, which addresses the scalability challenge of end-to-end inference for on-chip photonic neural networks. Nonlinear activation functions based on opto-electro-optical (O-E-O) conversion achieve positive net gain, overcoming network depth limitations. Through in-situ training, end-to-end inference is realized for image binary classification (96% accuracy) and handwritten digit quadri-classification (94% accuracy) tasks.

In recent years, graph-structured data has become increasingly prevalent across domains such as social networks and transportation systems. Efficient analysis of such data is critical for advancing intelligent applications across a wide range of fields. However, traditional network architectures, such as CNNs and FCNNs, struggle to effectively process graph-structured data. This limitation has led to the emergence of GNNs, which are specifically designed to process graph-structured data. GNNs perform computations by aggregating and transforming the local features of neighboring nodes, enabling efficient modeling of data dependencies while reducing redundant computations. Through message-passing mechanisms, GNNs iteratively fuse and update a node's intrinsic features with those of its neighboring nodes. This enables effective learning of complex topological structures and semantic relationships within graphs, delivering robust representational capabilities and generalization performance. Consequently, GNNs have become one of the most prominent architectures for graph representation and learning tasks. As the demand for real-time performance and energy efficiency increases, GNNs with low latency and intrinsic parallelism demonstrate significant application potential in social networks, traffic prediction, medical imaging, and intelligent decision-making, particularly in the processing of dynamic temporal information. Consequently, the third generation of neural network models—spiking neural networks (SNNs)—has attracted increasing attention due to their biological inspiration. Their core feature lies in transmitting information through discrete spikes. Unlike traditional neural networks, SNNs introduce a temporal dimension that simulates the dynamic firing behavior of biological neurons, thereby achieving greater functional and structural resemblance to biological neural systems. In particular, PSNNs have emerged as a promising research direction. PSNNs emulate the spike-firing mechanism of biological neurons by using light pulses as information carriers. Through the integration–threshold–firing process in photonic neurons (e.g., semiconductor lasers and silicon photonic devices) and weight modulation via photonic synapses, PSNNs achieve ultra-high speed, low-power, and intrinsic parallel computation. These models have been experimentally validated on small-scale image recognition tasks (e.g., MNIST handwritten digit classification)[17-20]. Research efforts are now advancing toward large-scale integration and adaptation to more complex application scenarios (e.g., edge intelligence)[21-22]. However, the integration of GNNs and PSNNs remains largely unexplored. Combining these two paradigms holds great potential for biologically inspired, energy-efficient processing of graph-structured data and interpretable reasoning, thus warranting further investigation.

Accordingly, this work proposes a novel PSGNN architecture that integrates photonic computing, SNNs, and GNNs:

1. This architecture integrates the efficient modeling capabilities of GNNs for non-Euclidean structured data, the temporal representation dynamics of spiking neurons, and the ultra-high speed and low-power advantages of photonic chips for parallel computation.

2. In the hardware–software co-optimization process, a bias-term simulation mechanism is constructed through a feature-dimension expansion strategy to satisfy the dual constraints of unbiased estimation and weight non-negativity, thereby enabling effective model training under these conditions. Classification experiments on the KarateClub (PubMed) datasets demonstrate that the proposed architecture achieves a training accuracy of 100% (92±2%) and a test accuracy of 97% (90±1%).

3. Classification experiments on the KarateClub dataset were conducted using a hardware verification platform constructed with a 4×4 Mach-Zehnder interferometer (MZI) array. The experimental results demonstrate that the proposed architecture achieves 93% testing accuracy on this dataset.

4. The architecture achieves an inference latency of 97 ps, demonstrating significant acceleration within the optical domain. In photonic linear computation, the system achieved an energy efficiency of 280 GOPS/W and a computational density of 52 GOPS/mm$^2$, indicating superior latency characteristics and high optoelectronic co-computational efficiency.

The experimental results demonstrate that the proposed PSGNN exhibits strong cross-domain applicability, supporting applications in biological protein structure prediction[23-25], dynamic topology analysis of social networks[26-27], intelligent traffic scheduling and early warning[27-28], and industrial IoT fault diagnosis, thereby contributing to real-time intelligent decision-making.

## 2 Principles and Methods

## 2.1 Photonic spiking graph neural network architecture

The PSGNN architecture, based on a silicon-photonics-based 4×4 MZI grid chip, is shown in Fig. 1(a). Its implementation is demonstrated through the KarateClub social network grouping analysis task. The KarateClub social network models community fragmentation scenarios caused by opinion divergence in real-world social groups. It contains 34 nodes, representing club members, and 78 edges, representing the social relationships among members, thereby characterizing the social structure. Based on this social network, graph-structured data are constructed by generating a feature matrix containing node-specific attributes and an adjacency matrix describing node connections. Both matrices serve as input to the GNN. The PSGNN comprises a three-layer structure: input layer Layer1, hidden layer Layer2, and output layer Layer3.

As shown in Fig. 1(a), the input graph-structured data are first fed into Layer 1 for preliminary processing of node features and their relationships. The data then propagate through the 4×4 MZI grid chip module in the hidden layer (Layer 2), which exploits its photonic properties to perform further computation and feature transformation. Subsequently, the data are transmitted to the output layer (Layer 3), where node feature representations are optimized, and classification results are produced, categorizing members into two groups: "Mr. Hi" and "Officer". The final classification results, shown in Fig. 1(b), demonstrate the distribution of members assigned to the "Mr. Hi" and "Officer" categories.

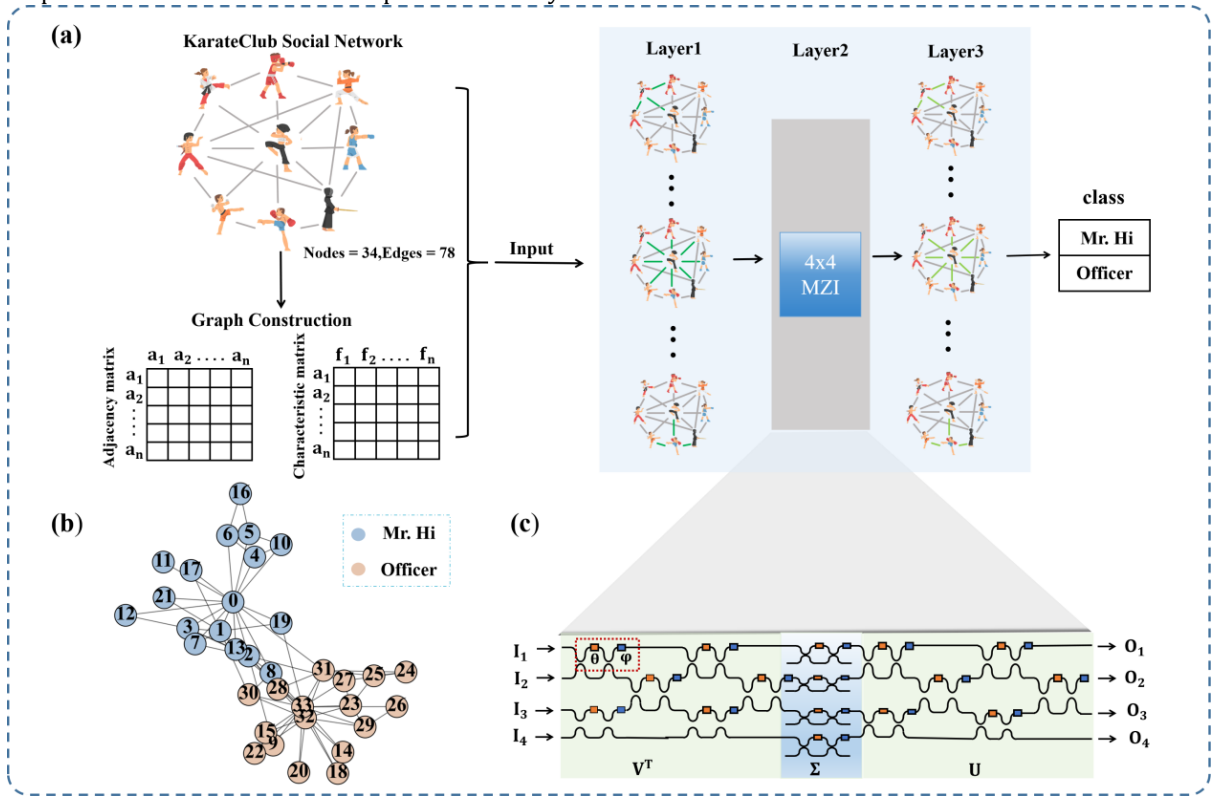

**Fig. 1.** Photonic spiking graph neural network. (a)KarateClub social network and graph structure construction. (b)Social network node classification. (c)Schematic diagram of the silicon-photonics-based 4×4 MZI array.

The detailed connection structure of the MZI rectangular architecture in Layer 2 is shown in Fig. 1(c). This architecture processes input signals, such as optical signals converted from pattern structure data, based on optical principles. The system performs core operations, including linear matrix multiplication and addition, through the coordinated operation of its optical components. The corresponding mathematical model is expressed as follows[29]:

$$O = WI = U\Sigma V^T \cdot I \quad (1)$$
$$W = U\Sigma V^T \quad (2)$$

Here, W denotes the weight matrix regulated by the phase shifter voltage, I ($I_1, I_2, I_3, I_4$) represents the input vector, O ($O_1, O_2, O_3, O_4$) denotes the output vector, $V^T$ and U correspond to matrix transformation modules, and Σ represents the summation module.

## 2.2 Working principle of spiking graph neural network

GNNs employ graph structures as the fundamental framework governing their propagation processes. This characteristic fundamentally differentiates them from conventional SNNs. Traditional SNN models lack the capability to represent and aggregate irregular topological connections[30], whereas GNNs can effectively integrate topological relationships and node features within graph structures[31].

Therefore, we propose PSGNN, as shown in Fig. 2, which aims to enable efficient processing of graph-structured data and biologically interpretable reasoning in a biologically inspired and energy-efficient manner. When integrating SNN models with graph structures, it is essential to address the limitation of treating individual nodes merely as independent SNN inputs. Instead, the state information of each node's neighborhood must be fully considered to comprehensively characterize the associative properties of nodes within the global graph topology. This approach prevents the loss of structural information that would otherwise occur when focusing solely on isolated nodes. The graph structure[32] is defined as G = (V, A) , where: V denotes the node set, i.e., V = {$v_1$, $v_2$, ..., $v_N$}, where N is the total number of nodes in the graph; $A \in \mathbb{R}^{N \times N}$ represents the adjacency matrix, describing the connection relationships and edge weights between nodes, with matrix element $a_{ij}$ corresponding to the edge weight value between nodes $v_i$ and $v_j$. The feature matrix is defined as $X \epsilon \mathbb{R}^{N \times d}$, containing the features vectors of all nodes in the graph, i.e., $X = [f_1, f_2, ..., f_N]$, where d denotes the dimensionality of each node feature vector. Additionally, define the degree matrix $D = diag(d_1, d_2, ..., d_N)$, where the diagonal elements $d_i$ represent the cumulative sum of the i-th row elements in the adjacency matrix A, i.e., $d_i = \sum_j a_{ij}$. This quantifies the connectivity degree of each node.

During the data input phase, the model receives two primary matrices as input. The first is the adjacency matrix A, which characterizes the connections between nodes in the graph. Specifically, if $a_{ij} = 0$, no direct edge exists between the two nodes, whereas if $a_{ij} = 1$, a direct edge is present. The normalized adjacency matrix input to the model is denoted as $\tilde{A}$(33):

$$\tilde{A} = A + \lambda I_N \quad (3)$$

Among these, the A-neighbor feature and the I-self feature are designed to ensure that each node incorporates its intrinsic characteristics during the information aggregation process. The feature matrix X serves as the self-feature information of each node.

During the data processing phase, the network captures feature correlations and intrinsic patterns among nodes through its embedded network layers. The information propagation process is expressed as follows[33]:

$$H^{(l+1)} = \sigma(\hat{D}^{-1}\hat{A}H^{(l)}W^l) \quad (4)$$

Here, $H^{(l)}$ denotes the node feature matrix at layer $l$, $\hat{D}$ represents the degree matrix corresponding to $\hat{A}$; and $W^l$ denotes the learnable parameter matrix.

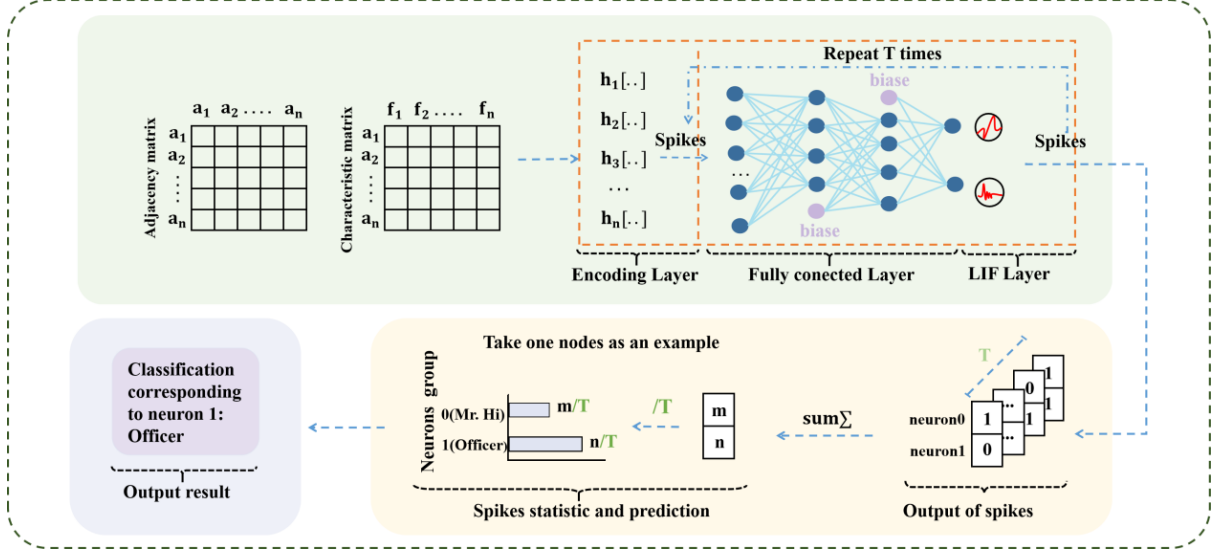

**Fig. 2.** Schematic of the photonic spiking graph neural network for social network clustering analysis

To simultaneously satisfy the dual constraints of unbiased estimation and non-negativity of weights during training, a bias-term simulation mechanism is introduced through feature-dimension expansion. Specifically, at each feature-input or transformation stage, an additional fixed-dimension "bias channel" is appended to the current feature matrix. This bias channel has two core properties. First, all elements within the channel are set to 1. Accordingly, for a feature matrix $X \in \mathbb{R}^{N \times d}$, where N denotes the number of samples and d the original feature dimension, the expanded feature matrix is represented as $X' \epsilon \mathbb{R}^{N \times (d+1)}$, with an additional column vector $1 \in \mathbb{R}^N$, where $1_i = 1$ for all $i \in [1, n]$; Second, an independent learnable parameter $w_{biase} \epsilon \mathbb{R}_{\geq 0}$ is assigned to this bias channel. This parameter is incorporated into the non-negativity constraint optimization alongside with the weight parameters associated with the original feature dimensions.

Finally, the system outputs two-dimensional feature representations, which are then used to categorize the nodes in the social network into two distinct groups, thus completing the clustering analysis task.

### 2.3 4×4 MZI chip and experimental setup

A hardware experimental platform was developed using a silicon-photonics-based 4×4 MZI array, with the MZI chip acting as the core computational unit[32-33]. The system is designed around the technical workflow of optical signal loading, on-chip linear processing, electro-optical conversion, electrical signal demodulation and analysis. It integrates a light source, optical modulator, control module, and detection unit to form a closed-loop experimental platform. The overall system architecture and signal flow are shown in Fig. 3(a).

The photonic computing system utilizes four tunable laser light sources to provide parallel optical signal inputs to the chip. In the signal modulation stage, a Mach–Zehnder modulator (MZM) receives electrical signals from the field-programmable gate array (FPGA) through a digital-to-analog converter (DAC) and linearly maps the digital input data into optical intensity signals. A polarization controller (PC) is connected in series between the light source and the chip. By adjusting the PC, the output optical power is maximized, ensuring that the optical signal enters the chip waveguide with optimal coupling efficiency.

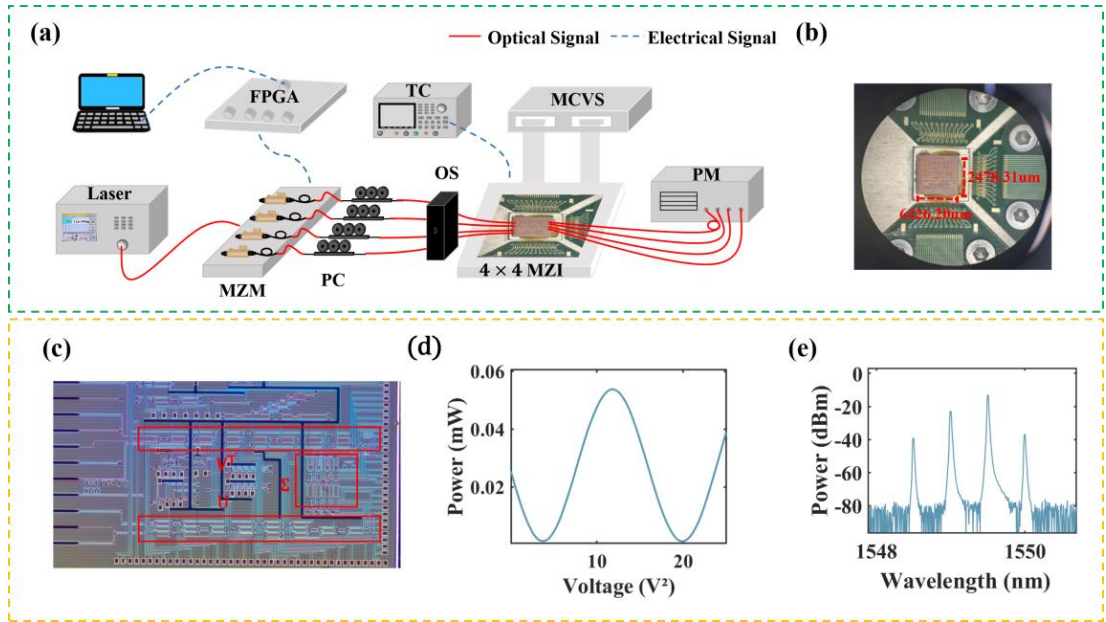

**Fig. 3.** Experimental setup and chip test results. (a)Experimental system structure and signal flow. (b) 4×4 MZI chip fabricated on an insulator-on-silicon platform. (c) 4×4 rectangular architecture. (d)On-chip phase shifter diagram. (e)MZI spectral response.

The chip control module integrates a multi-channel voltage source and a temperature controller (TC). The detection and analysis module includes an optical power meter and an oscilloscope. During the training phase, the optical power meter measures the four optical power outputs from the chip for matrix similarity evaluation. In the verification phase, the oscilloscope captures the electrical signals demodulated from the MZM, enabling error analysis by comparing them with the theoretical algorithmic outputs. Additionally, an optical switch (OS) enables independent switching among the four optical inputs. Working in coordination with the optical power meter, it allows stepwise acquisition of the column vectors of the chip's transmission matrix.

The 4×4 rectangular MZI chip architecture, shown in Fig. 3(c), consists of sixteen 2×2 MZI basic units interconnected through cascading and cross-connections. Each 2×2 MZI unit integrates four electro-optical phase shifters—two located in the inner arms and two in the outer arms.

The 4×4 MZI chip is fabricated on an insulator-on-silicon platform, as shown in Fig. 3(b). Its core functionality is to perform linear matrix multiply–accumulate (MAC) operations, which can be mathematically expressed as $O = W \cdot I = U\Sigma V^T \cdot I$. Using the algorithmically predefined target weight matrix Wtarget = $U\Sigma V^T$ (of dimension 4×4, with element values ranging from 0 to 1) as the reference, the chip is trained using the Stochastic Parallel Gradient Descent (SPGD) algorithm[34]. The detailed implementation logic of this training process can be understood by referring to the flowchart shown in Fig. 4, together with the step-by-step explanation provided below.

Step1-Initialization: Set the initial values of control parameters, iteration counters, and related variables.

Step2-Random Perturbation: Apply random perturbations to the control parameters to generate new combinations.

Step3-Performance Evaluation: Evaluate the performance metrics corresponding to the newly generated parameter sets.

Step4-Gradient Estimation: Estimate the gradients of the control parameters based on the variations in performance metrics.

Step5-Parameter Update: Update the control parameters using the estimated gradient information.

Step6-Iteration: Repeat Steps 2–5 until the termination condition is satisfied.

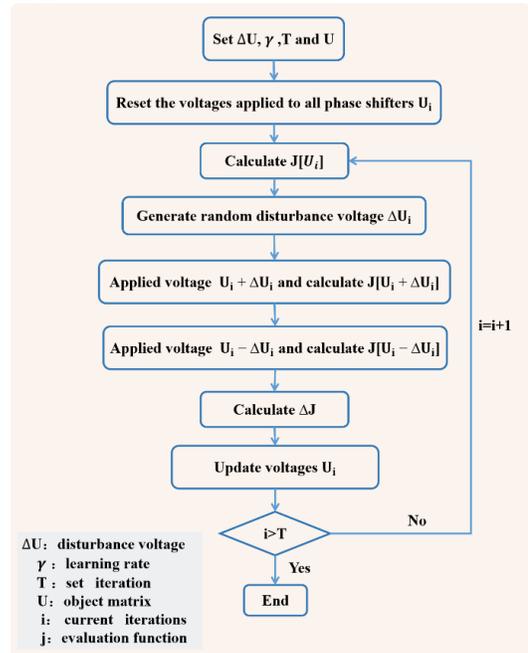

**Fig. 4.** Flowchart of the stochastic parallel gradient descent algorithm

During the training process, the voltages of the phase shifters are dynamically adjusted through a multi-channel voltage source. Every ten iterations, the similarity between the current transmission matrix $W_{current}$ and the target matrix $W_{target}$ is computed. The similarity metric is calculated using the following formula[35]:

$$corr = \frac{dot(W_{current}, W_{target})}{\sqrt{dot(W_{current}, W_{target}) \cdot dot(W_{target}, W_{target})}} \quad (5)$$

Here, dot($\cdot,\cdot$) denotes the matrix dot product operation.

The on-chip phase shifter, shown in Fig. 3(d), is designed based on the silicon electro-optic effect and employs a doped silicon waveguide structure. In the experiments, each phase shifter is independently controlled through a multi-channel voltage source and plays a key role in aligning the chip's transmission matrix with the target weight matrix. The control precision directly determines the upper bound of convergence for matrix similarity. The spectral response of the MZI is shown in Fig. 3(e), where the spectrum at each output port represents the linear superposition of four output wavelengths.

## 3 Experimental Results

### 3.1 4×4 MZI experimental results

In the photonic hardware inference experiment, we used the 4 × 136 data points output from Layer 1 as input for the hardware experiment. Layer 2, the network's hidden layer, was implemented on the photonic neuromorphic chip. The trained and optimized phase shifter voltage parameters were loaded onto the chip for processing. The FPGA transmitted the task input vector I, which was modulated onto light via the MZM. Linear multiply-accumulate operations were performed using the MZI chip, generating optical signals. These signals were converted into electrical signals via the PD, transmitted to an oscilloscope for acquisition, and compared against the theoretical algorithm output $O_{target} = W_{target} \cdot I$. The relative error was calculated as follows[36]:

$$\delta = \frac{|O_{measured} - O_{theory}|}{O_{theory}} \times 100\% \quad (6)$$

Here, $O_{measured}$ represents the experimentally obtained result, while $O_{theory}$ denotes the theoretical ideal output.

The algorithm input consists of a feature dataset containing four channels, as shown in Fig. 5(a), with a total length of 544 data points corresponding to the sequence obtained by row-wise flattening of the 4×136 matrix. Among these, Raw_Input represents the data output from Layer 1, Act_Input denotes the activated data derived from the Layer 1 output, and Act_Output indicates the theoretical target result obtained after processing the data through Layer 2. Representative results from the MZI linear computation experiment for one channel were selected and compared with the theoretical target outputs, as shown in Fig. 5(c). As shown, the actual output from the MZI linear computation closely matches the target output computed by the algorithm, with a relative error approaching 0%.

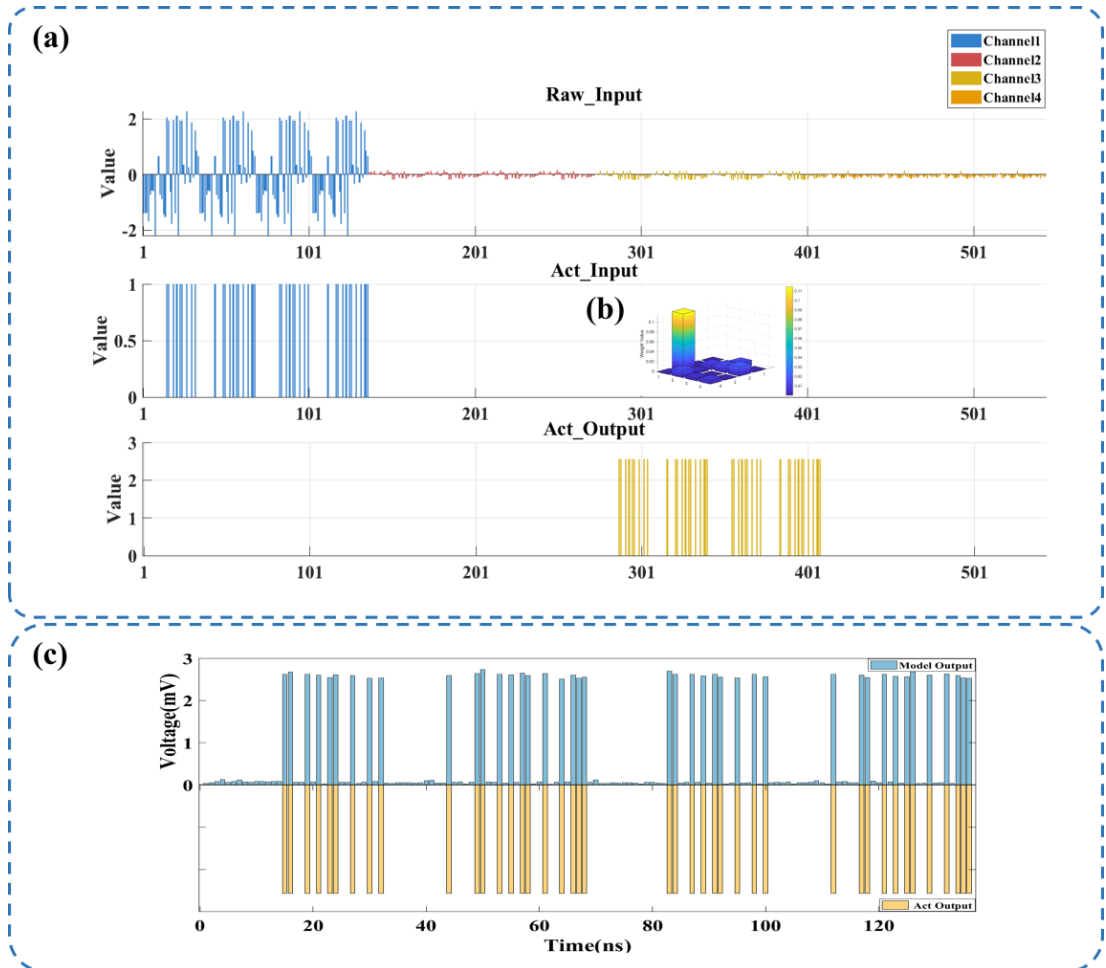

**Fig. 5.** Theoretical input–output relationship, weight parameters, and experimental validation of the algorithm. (a) Theoretical input–output relationship. (b) Weight parameters. (c) Comparison between the experimental linear output and the theoretical output.

## 3.2 Experimental results on software-hardware co-perception optimization in KarateClub

Fig. 6(a) shows the comparison of loss and accuracy between the algorithmic simulation and the photonic computing experiments. From the loss curve, the algorithm's training loss decreases rapidly during the initial training phase and gradually stabilizes thereafter. In contrast, the overall reduction in training loss observed in the photonic experiment is relatively modest and consistently remains higher than that of the algorithmic simulation. Analysis of the accuracy curve reveals that the accuracies obtained from both algorithmic training and photonic hardware training gradually converge to 1 after sufficient iterations. This result indicates that the accuracy performance of both approaches demonstrates a high degree of consistency upon training convergence.

The test results of the algorithmic simulation and the photonic computing experiments are presented in Table 1. The comparison is performed across four evaluation metrics: precision, recall, F1-score, and overall accuracy.

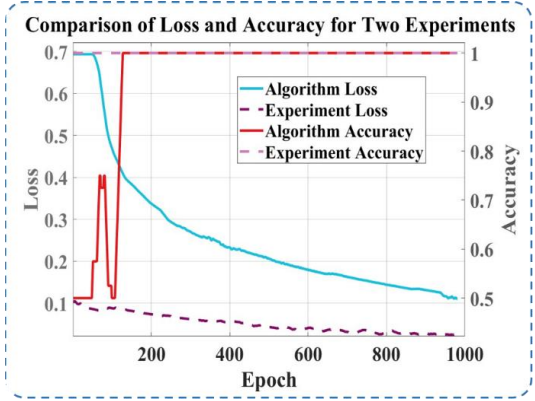

**Fig. 6.** Comparison of training and testing performance between the algorithmic simulation and the photonic hardware experiment

**Table 1.** Test results of the algorithmic simulation and the photonic computing experiments

**Aigorithm(Experiment) Test Results**

|  | Percision | Recall | f1-score |
|---|---|---|---|
| Mr.Hi | 1.00(1.00) | 0.94(0.82) | 0.97(0.90) |
| Officer | 0.94(0.85) | 1.00(1.00) | 0.97(0.92) |
| Accuracy | 0.97(0.93) | 0.97(0.91) | 0.97(0.91) |

In the "Mr. Hi" category, the algorithm achieved precision, recall, and F1-score values of 1.00, 0.94, and 0.97, respectively, while the corresponding experimental results were 1.00, 0.82, and 0.90.

In the "Officer" category, the algorithm achieved precision, recall, and F1-score values of 0.94, 1.00, and 0.97, respectively, whereas the corresponding experimental results were 0.85, 1.00, and 0.92.

In terms of overall performance, the algorithm achieved an accuracy of 0.97 compared with 0.93 for the experiment, a recall of 0.97 versus 0.91, and an F1-score of 0.97 against 0.91.

According to the experimental results, photonic computing exhibited slightly lower task performance than the algorithmic approach. However, in terms of inference efficiency, photonic computing achieved an inference latency of 97 ps, substantially outperforming conventional electronic computing systems. Considering both task performance and inference efficiency, the experimental results demonstrate the superior overall practicality of photonic computing, making it particularly suitable for application scenarios requiring ultralow latency.

## 3.3 Experimental results on software-hardware co-perception optimization in PubMed

The PubMed dataset consists of 19,717 biomedical literature nodes and 44,324 undirected edges, constructed based on citation relationships. Each node feature is represented by a 500-dimensional TF–IDF vector. After being processed by a graph neural network model, the nodes are classified into three categories corresponding to diabetes-related topics. The detailed experimental results are presented in Table 2.

**Table 2.** Training and testing accuracy across different network scales and time steps (T) on the PubMed dataset

| Ablation Study | Parameter | Train Accuracy | Test Accuracy |
|---|---|---|---|
| T | 1 | 89% | 88.92% |
|  | 2 | 91% | 88.87% |
|  | 4 | 90% | 88.59% |
|  | 6 | 91% | 89.16% |
| Scale | 8x8 | 89% | 88.06% |
|  | 16x16 | 91% | 89.16% |
|  | 32x32 | 93% | 90.23% |
|  | 64x64 | 94% | 91.03% |

In the ablation study with the time-step parameter T, the training accuracy was 89%, and the testing accuracy was 88.92% when T=1. When T=2 and T=6, the training accuracy in both cases was 91%, with corresponding testing accuracies of 88.87% and 89.16%, respectively. For T=4, the training and testing accuracies were 90% and 88.50%, respectively.

When the network size was 8×8, the training and testing accuracies were 89% and 88.06%, respectively. At 16×16, the training and testing accuracies were 91% and 89.16%, respectively. At 32×32, the training and testing accuracies were 93% and 90.23%, respectively. At 64×64, the training and testing accuracies were 94% and 91.03%, respectively.

The above results provide quantitative evidence supporting the further optimization of the model's performance in node classification tasks with this type of structured data.

## 4 Discussion

In this section, we examine the effects of network size and time step on the experimental results. As shown in Fig. 7(a)–(d), ablation experiments for algorithms with different network sizes (H = 2, 3, 4, and 7) show how training loss and accuracy evolve over iterations. As H increases, model training converges more rapidly, the final loss decreases, and the accuracy improves in a faster and more stable manner. A larger network size enhances the model's fitting capability and training efficiency.

The effect of different T values (set to T = 1, 2, 6, and 8) on the model training process is shown in Fig. 7(e)–(h). For all T values, the training loss continuously decreases as the training iterations increase, while the training accuracy steadily increases. Regardless of the specific T value, the model effectively learns from the training data throughout the entire training process. Analysis of training performance across different T values indicates that an appropriately chosen T enables the network to focus on the most salient aspects of the input representation for decision-making, significantly improving the stability and efficiency of the training process.

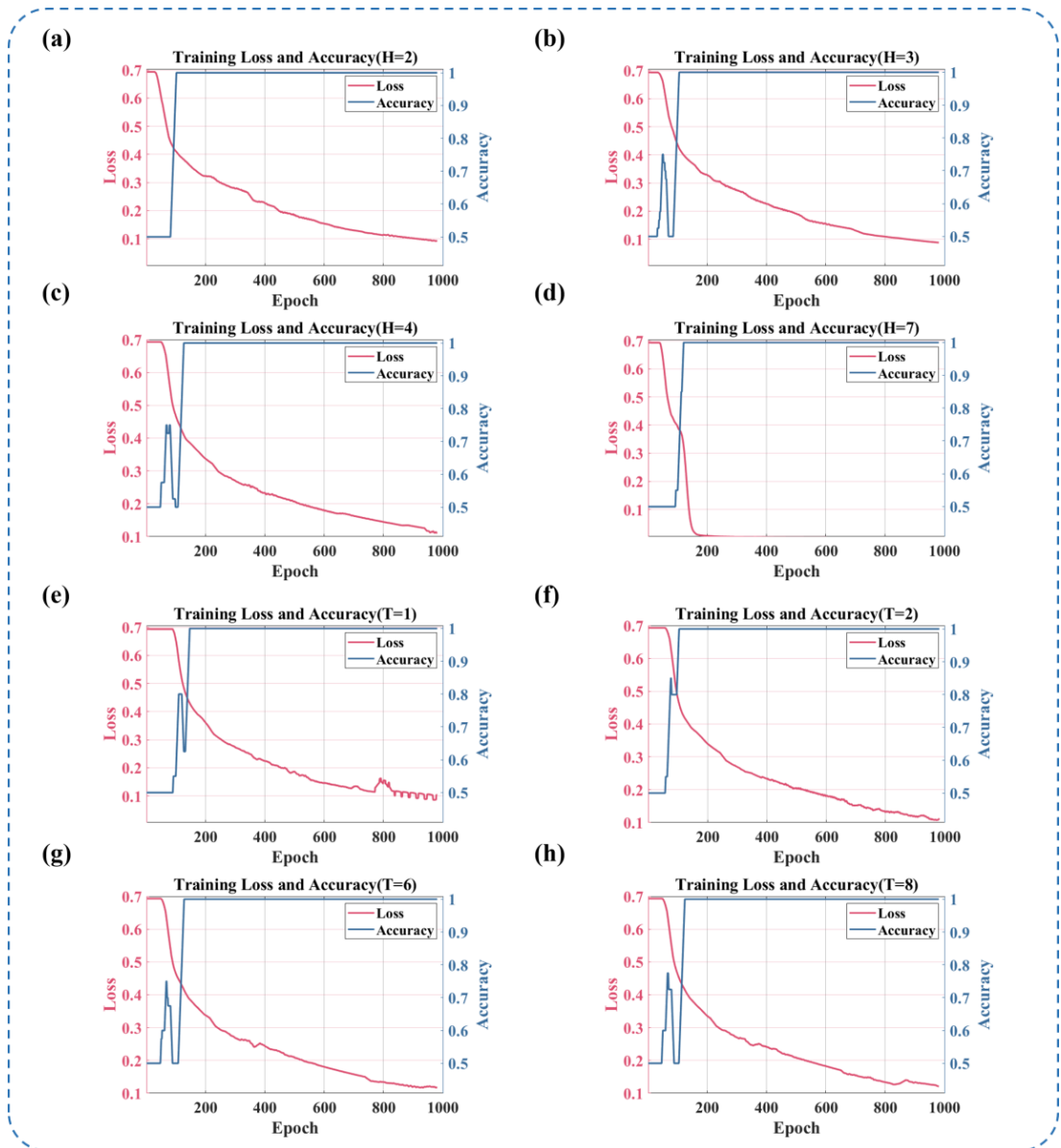

**Fig. 7.** Ablation study for different network scales and time-step size

The initial experimental setup was as follows: two core nodes were designated as supervised nodes for training, while all other nodes in the network collectively served as the validation set. To further evaluate the robustness of the model's performance and the impact of key parameters, an ablation experiment was conducted. Building on the original experimental framework, the data were split into training and testing sets with a 7:3 ratio. The ablation experiment specifically investigated two combinations of key parameters: time steps and network sizes. The detailed results of this ablation experiment are presented in Table 3.

**Table 3. Train and test accuracy across different T and network scales**

| Abliation Study | Parameter | Train Accuracy | Test Accuracy |
|---|---|---|---|
| T | 1 | 34% | 14% |
|  | 2 | 100% | 100% |
|  | 4 | 100% | 100% |
|  | 6 | 100% | 100% |
| Scale | 2x2 | 96% | 100% |
|  | 4x4 | 100% | 100% |
|  | 5x5 | 100% | 100% |
|  | 7x7 | 100% | 100% |

To validate the robustness of the model's performance and assess the impact of key parameters on experimental outcomes, this study designed ablation experiments built on the initial experimental framework. The core differences between the two approaches lie in data processing logic and experimental focus. The initial experiments focused on the foundational model training and validation. The initial experiments did not separate the training and testing sets; instead, only two core nodes were selected as supervised nodes during training. Meanwhile, all network nodes collectively served as the validation set, establishing the basic experimental framework and performing preliminary model validity checks. The ablation experiments, serving as an advanced validation phase, adjust the data partitioning method. The experimental data are explicitly split into training and testing sets at a 7:3 ratio. The model's generalization capability is evaluated using the independent testing set to assess performance robustness. Furthermore, the experiments focus on analyzing key parameters by specifically designing test plans targeting two parameter combinations: "time step size" and "network size." This approach quantifies the specific impact of different parameter configurations on model performance.

## 5.Conclusion

This paper presents a PSGNN architecture. At the algorithmic level, the proposed architecture optimizes the spatiotemporal encoding mechanism of spiking signals to leverage the high-speed propagation and parallel interference characteristics inherent to photonic signals. At the hardware level, it adjusts the voltage–phase response range of the on-chip phase shifters and the optical modulation precision of the modulators to meet the computational accuracy requirements of the PSGNN, thereby effectively reducing latency losses during data conversion and computation. Through coordinated hardware-software co-optimization, efficient collaborative computation between the spiking graph neural network and the photonic neuromorphic chip is realized. After deploying the spiking graph neural network onto the photonic neuromorphic chip, the model achieved a test accuracy of 93%. These results demonstrate that the proposed PSGNN architecture exhibits strong robustness to hardware noise and errors, while maintaining stable computational performance under non-ideal hardware conditions. Therefore, to address the increasing demand for low-latency computing in domains such as social network analysis, recommendation systems, and biomedicine, we propose a high-speed, low-latency PSGNN architecture. This architecture holds promise as a core technological enabler in these critical domains, further promoting the deep integration of photonic computing chips, SNNs, and GNNs. It provides essential theoretical foundations and technical references for the development of next-generation low-power, and highly parallel intelligent computing systems.

## Data Availability

The data that support the findings of this study are available from the corresponding author upon reasonable request.

## Author information

### Corresponding author:

*Shuiying Xiang-State Key Laboratory of Integrated Service Networks, Xidian University, Xi'an 710071, China and State Key Discipline Laboratory of Wide Bandgap Semiconductor Technology, School of Microelectronics, Xidian University, Xi'an 710071, China.*
*Email: syxiang@xidian.edu.cn*
*Xingxing Guo- State Key Laboratory of Integrated Service Networks, Xidian University, Xi'an 710071, China and State Key Discipline Laboratory of Wide Bandgap Semiconductor Technology, School of Microelectronics, Xidian University, Xi'an 710071, China.*
*Email: xxguo@xidian.edu.cn*

### Authors:

*Wanting Yu-State Key Laboratory of Integrated Service Networks, Xidian University, Xi'an 710071, China.*
*Shuiying Xiang-State Key Laboratory of Integrated Service Networks, Xidian University, Xi'an 710071, China and State Key Discipline Laboratory of Wide Bandgap Semiconductor Technology, School of Microelectronics, Xidian University, Xi'an 710071, China.*
*Xingxing Guo- State Key Laboratory of Integrated Service Networks, Xidian University, Xi'an 710071, China and State Key Discipline Laboratory of Wide Bandgap Semiconductor Technology, School of Microelectronics, Xidian University, Xi'an 710071, China.*
*Shangxuan Shi-State Key Laboratory of Integrated Service Networks, Xidian University, Xi'an 710071, China.*
*Haowen Zhao-State Key Laboratory of Integrated Service Networks, Xidian University, Xi'an 710071, China.*
*Xintao Zeng-State Key Laboratory of Integrated Service Networks, Xidian University, Xi'an 710071, China.*
*Yahui Zhang-State Key Laboratory of Integrated Service Networks, Xidian University, Xi'an 710071, China.*
*Hongbo Jiang-State Key Laboratory of Integrated Service Networks, Xidian University, Xi'an 710071, China.*
*Yue Hao-State Key Discipline Laboratory of Wide Bandgap Semiconductor Technology, School of Microelectronics, Xidian University, Xi'an 710071, China.*

## Competing interests

The authors declare no competing financial interests.

## Funding

This work was supported by the National Natural Science Foundation of China(No.62535015, 62575231); The Fundamental Research Funds for the Centrale Universities (QTZX23041); Xidian University Specially Funded Project for Interdisciplinary Exploration (TZJH2024009)；The China Postdoctora Science Foundation (Certificate Number: 2024M752526).